\documentclass[preprint,showpacs,preprintnumbers,amsmath,amssymb]{revtex4}
\usepackage{graphicx}
\usepackage{dcolumn}
\usepackage{bm}
\begin{document}
\title{Elementary excitations of trapped Bose gas in the large-gas-parameter regime}
\author{Arup Banerjee and Manoranjan P. Singh}
\affiliation{Laser Physics Division, Centre for Advanced Technology, Indore 452013, India}
\date{24 June 2002}
\pagebreak
\begin{abstract}
We study the effect of going beyond the Gross-Pitaevskii theory on the frequencies of collective
oscillations of a trapped Bose gas in the large gas parameter regime. We go beyond the 
Gross-Pitaevskii regime by including a higher-order term in the interatomic correlation energy.
To calculate the frequencies we employ the sum-rule approach of many-body response theory coupled
with a variational method for the determination of ground-state properties.  We show that going 
beyond the Gross-Pitaevskii approximation introduces significant corrections
to the collective frequencies of the compressional mode.
\end{abstract}
\maketitle
\section{Introduction}
The mean-field Gross-Pitaevskii (GP) theory \cite{gp} has been
quite successful in explaining both static and dynamic properties
\cite{dalfovo,fetterrv,castinrv} of Bose-Einstein condensates (BEC)
produced in alkali atoms \cite{anderson,bradley, davis} . The reason for the success
of the GP theory is mainly the satisfaction of dilute gas condition
$na^{3}<<1$ ( where $n$ is atomic density, $a$ is the s-wave
scattering length of interatomic potential and the parameter
$na^{3}$ is called the gas parameter) in the above mentioned
experiments. Typically the values of gas parameter in
these experiments were in the range of $10^{-5}-10^{-4}$. The gas parameter can
have a large value when the number of atoms $N$ in the condensate or the scattering length
is large.
Thus in the later  experiments reported in Refs. \cite{ketterle,stamperkurn}
$N$ was of the order of $10^{8}$ and 
the corresponding value of gas parameter was $10^{-3}$. 
On the other hand, in a very
recent experiment \cite{cornish} the maximum value of the
peak gas parameter was $n(0)a^{3}\approx 10^{-2}$ ($n(0)$ is the
peak density of the condensate) although the condensate contained only
$10^{4}$ $^{85}$ Rb atoms. In this experiment the
scattering length $a$ could be varied continuously from a negative to a
very large positive value by exploiting the Fesbach
resonance.  These kind of
condensates, then, naturally raise the question about the validity
of the mean-field GP approach. Theoretical investigations of the effects of going
beyond the GP (BGP) theory have already been reported in the
literature \cite{timmermans,braaten,nunes,polls,banerjee}.
Recently \cite{fabrocini} ground-state properties of BEC with
large values of gas parameter (as achieved in the experiments \cite{cornish})
have been theoretically studied to assess the accuracy of the GP theory. It has
been shown that for large values of the gas parameter ($\approx$ $10^{-2}$)
the total energy and the chemical potential
are modified significantly by the BGP theory.
It is well known that these changes are also reflected in the
frequencies of collective excitations which can be
measured with a greater accuracy. Therefore,
by measuring the frequencies of collective excitations it should be possible to 
observe the effect of  the BGP theory.
This has motivated us to calculate the
corrections to the frequencies of collective oscillations due to the BGP theory.
We focus our attention on the condensates with values of
gas parameters, spanning a range similar to that achieved in the experiment \cite{cornish}
and make an assessment of the accuracy of the GP theory in
predicting the frequencies of collective oscillations of such condensates.

In the past corrections to the frequencies of collective
oscillations of trapped bosons arising due to the BGP theory have been estimated
\cite{pitaevskii,braaten2} within the Thomas-Fermi (TF)
approximation which is valid in the large $N$ limit (more precisely when $Na/a_{h0}>>1$ is
satisfied, where $a_{h0}$ is the characteristic length of trapping potential and it is defined
in the next section). In the aforesaid references corrections to the frequencies
were obtained by solving the time-dependent second-order differential equation perturbatively for
the density. Thus the analytical expressions for the fractional change in the frequencies 
are correct only up to the first order in $\sqrt{n(0)a^{3}}$.
It is, then, expected that these expressions for the corrections
will not be accurate in the large-gas-parameter regime, as achieved in the
recent experiment \cite{cornish}.
In this paper we also derive an analytical expression for the correction within the TF
approximation by making use of the sum-rule approach. We find that the expression for the
correction obtained by us is valid for somewhat larger range of gas parameters than those of
Refs. \cite{pitaevskii,braaten2}. As expected, for very small gas parameters the expression 
derived by us correctly reduces
to those of Refs. \cite{pitaevskii,braaten2}.
We will return to the question of validity of these analytical results later.

It is  well known that the TF approximation does
not reproduce the surface region of density profile of trapped
condensates accurately \cite{dalfovo}. On the other hand, the collective
oscillation frequencies characterizing the response properties of
the trapped gas crucially depend on the tail or the surface
region of the density profile \cite{bertsch}.  Therefore, it becomes necessary
to go beyond the TF approximation especially for the calculation
of frequencies of collective oscillations \cite{braaten2}.
To accomplish this task we make use
of well established variational method
\cite{banerjee,byam,fetter1,mp} to determine the ground-state
densities. The collective oscillation frequencies are then obtained by using the 
sum-rule approach \cite{bohigas,lipparini} of
many-body response theory. In the sum-rule approach the
calculation of frequencies requires the knowledge of the ground-state wave function
(or the ground-state density) of the many-body system only.
Therefore the accuracy of frequencies determined by the sum-rule
approach depends on the accuracy of the  ground-state
wave function (ground-state density) employed for the calculations. 
Keeping this in mind our aim is also to obtain
reasonably accurate ground-state densities of the trapped
condensates. To this end we employ variational approach by using an energy functional of $N$
interacting trapped bosons within the local density approximation
(LDA) along with a judiciously chosen variational ansatz for the
ground-state density \cite{banerjee,mp}.

The paper is organized in the following manner. In Section II we
describe the theoretical method for the calculation of the
frequencies of collective oscillations. Section III contains the analytical estimate of
the corrections to the frequencies within the TF approximation, and
results of our variational calculation. The
paper is concluded in Section IV.

\section{Variational Sum-rule Approach}
The central quantity in the variational approach is the energy
functional for a condensate of $N$ bosons, each with mass $m$, confined in a trap potential
$V_{t}(\bf r)$:
\begin{equation}
E[n] = \int d{\bf r}\left [- \frac{\hbar^{2}}{2m}\sqrt{n({\bf r})}{\bf \nabla}^{2}
\sqrt{n({\bf r})} +
V_{t}({\bf r})n({\bf r}) + \epsilon(n)n({\bf r}) \right ],
\label{1}
\end{equation}
where, the first, the second, and the third terms
represent the kinetic energy of bosons, the energy due to the
trapping potential, and the energy due to interatomic correlation
within the LDA, respectively. Following the usual practice
we use the perturbative expansion for $\epsilon(n)$
in terms of the gas parameter $na^{3}$,
\begin{equation}
\epsilon(n) = \frac{2\pi\hbar^{2}an}{m}\left [1 + \frac{128}{15\sqrt{\pi}}\left (na^{3}
\right )^{\frac{1}{2}} + 8\left (\frac{4\pi}{3} - \sqrt{3}\right )(na^{3})\ln\left (na^{3}\right )
+ {\cal O}\left(na^{3}\right)\right ].
\label{2}
\end{equation}
The first term in the above expansion, which corresponds to the
energy of the homogenous Bose gas within the mean-field theory as
considered in the GP theory, was calculated by Bogoliubov
\cite{bogoliubov}. The second term was obtained by Lee, Huang and
Yang (LHY) \cite{lee}, while the third term was first calculated
by Wu \cite{wu} using the hard-sphere model for the interatomic
potential. Although, it has been emphasized in the literature that the above expansion
is valid only for $na^{3}<<1$, it is only recently that the range of validity of above expansion
has been systematically
investigated by Giorgini et al. \cite{giorgini}. They have used diffusion
Monte Carlo (DMC) method to calculate the ground-state of uniform
gas of bosons interacting through different model potentials with
$a>0$. It has been found that the expansion (\ref{2}) is valid as
long as $na^{3}<10^{-3}$. However, for the values of the gas parameter beyond
$10^{-3}$ the inclusion of the logarithmic term leads to severe
mismatch with the DMC simulation results. On the other hand, expansion (\ref{2})
up to the LHY term gives an accurate representation of the DMC calculations even for the gas
parameter of the order of $10^{-2}$.  Consequently, we do not
consider the logarithmic term in the expansion (\ref{2}) 
for all the calculations in this study.

The trapping potential $V_{t}({\bf r})$ is taken to be
axially symmetric
characterized by two angular frequencies $\omega_{\bot}^{0}$ and $\omega_{z}^{0}$
($\omega_{x}^{0}=\omega_{y}^{0}=\omega_{\bot}^{0}\neq\omega_{z}^{0}$).
It is given by
\begin{equation}
V_{t}({\bf r}) = \frac{m{\omega_{\bot}^{0}}^{2}}{2}\left (x^{2} +
y^{2} + \lambda_{0}^{2}z^{2} \right ), \label{3}
\end{equation}
where $\lambda_{0} = \omega_{z}^{0}/\omega_{\bot}^{0}$ is the anisotropy parameter of
the trapping potential ($\lambda_{0}=1$ corresponds to a spherically symmetric trap).

The ground-state properties which are needed to calculate the
collective frequencies are determined by minimizing the energy
functional (\ref{1}) with respect to the density $n({\bf r}) 
$\cite{hk} satisfying the normalization condition
\begin{equation}
\int n({\bf r})d{\bf r} = N.
\label{norm}
\end{equation}
To implement the variational scheme we take the ansatz for the ground-state density as
\cite{banerjee,mp}:
\begin{equation}
n({\bf r}_{1}) = \frac{p}{2 \pi \Gamma(\frac{3}{2 p})}
\lambda^{\frac{1}{2}} \left(\frac{\omega_{\bot}}{\omega_{\bot}^0}\right)
^{\frac{3}{2}} e^{-
 \left(\frac{\omega_{\bot}}{\omega_{\bot}^0}\right)^p
\left( r_{1 \bot}^2 +\lambda z_1^2\right)^p},
\label{4}
\end{equation}
where $\lambda$, $\omega_{\bot}$ and $p$ are the variational
parameters which are obtained by minimizing $E[n]$ with respect to
these parameters. In the above equation we have used the scaled
co-ordinates ${\bf r}_{1}={\bf r}/a_{h0}$ with $a_{ho} = \left
(\hbar/m\omega_{\bot}^{o}\right )^{\frac{1}{2}}$. We note that
$n({\bf r_{1})}$ is normalized to unity and from Eq. (\ref{norm})
it can be seen that
\begin{equation}
n({\bf r}) = \frac{N}{a_{h0}^{3}}n({\bf r_{1})}.
\end{equation}
We make this choice of variational form for the density as it has
been shown convincingly that this ansatz is capable of describing
the ground state of Bose gas in a trap quite accurately both
within \cite{mp} and beyond \cite{banerjee} the GP
theory for a wide range of particle numbers. Using this
variational form the physical observables can be expressed
analytically in terms of three variational parameters
\cite{banerjee}. The total energy $E_{1} =
E/\hbar\omega_{\bot}^{o}$ can be written as \cite{banerjee}
\begin{equation}
\frac{E_{1}}{N} = T + U + E^{1}_{int} + E^{2}_{int}.
\label{5}
\end{equation}
Here $T$ and $U$ denote the average kinetic and the trapping
potential energies per particle, respectively. The term
E$^{1}_{int}$ gives the interaction energy per particle in the
mean-field approximation as considered in the GP theory while
E$^{2}_{int}$ gives the correction in the interaction energy due to the LHY term in the
expansion (\ref{2}) arising from the BGP theory.

The ground-state energy components obtained from the variational calculation are 
employed to compute the frequencies of collective oscillations by
using the sum-rule approach of many-body response theory. In the following we briefly outline the
sum-rule approach and present some of the results which are
relevant for the present work.

According to the basic result of the sum-rule approach \cite{bohigas,lipparini} the upper bound of the
lowest excitation energy is given by
\begin{equation}
\hbar\Omega_{ex} = \sqrt{\frac{m_{3}}{m_{1}}},
\label{6}
\end{equation}
where
\begin{equation}
m_{p} = \sum_{n}|\langle 0|F|n\rangle |^{2}\left (\hbar\omega_{n0}\right )^{p},
\end{equation}
is the $p$-th order moment of the excitation energy associated with the
excitation operator $F$ and $\Omega_{ex}$ is the frequency of excitation.
Here $\hbar\omega_{n0}=E_{n}-E_{0}$ is
the excitation energy of eigenstate $|n\rangle$ of the Hamiltonian
$H$ of the system. The upper bound given by Eq. (\ref{6}) is close to the exact
lowest excited state when this state is highly collective, that
is, when the oscillator strength is almost exhausted by a single
mode. This condition is satisfied by the trapped bosons in most of
the cases. Moreover, Eq. (\ref{6}) can be used for computation of
the excitation energies by exploiting the fact that the moments
$m_{1}$ and $m_{3}$ can be expressed as expectation values of
the commutators between $F$ and $H$ in the ground state
$|0\rangle$ \cite{bohigas,lipparini}:
\begin{eqnarray}
m_{1} & = & \frac{1}{2}\langle 0|\left [F^{\dagger},\left [H,F\right ]\right ]|0\rangle , \nonumber \\
m_{3} & = & \frac{1}{2}\langle 0|\left [\left [F^{\dagger},H\right ],\left [\left [H,[H,F\right ]\right ]\right ]
|0\rangle .
\label{7}
\end{eqnarray}
The main advantage of the sum-rule approach is that it allows us
to calculate the dynamic properties like excitation frequencies of
many-body systems with the knowledge of ground state $|0\rangle$
(or the ground-state density) only. Thus, with the knowledge of a reasonably
accurate ground-state density we calculate the expectation values
given in Eq.(\ref{7}) to determine the frequencies of collective
oscillations. It is worth mentioning that sum-rule approach has
been extensively used to calculate collective frequencies of the
trapped condensate \cite{stringari,kimura}.

For the purpose of calculation one needs to choose an appropriate
excitation operator $F$. In this paper we
concentrate only on the compressional collective mode
characterized by the $z$-component of angular momentum index, $m^{\prime}=0$.
For the axially symmetric trap the mode $m^{\prime}=0$ involves coupling of the
monopole ($l=0$) and the quadrupole ($l=2$) modes. Following
Kimura et al. \cite{kimura} the operator $F$ for the excitation of
$m^{\prime}=0$ collective mode is written as
\begin{equation}
F = \sum_{i}\left (x_{i}^{2} + y_{i}^{2} - \alpha z_{i}^{2}\right ),
\label{8}
\end{equation}
where $\alpha$ is a parameter characterizing the coupling of the two
modes due to the axial symmetry of the trap. This parameter is
obtained by making the excitation energy given by
Eq.(\ref{6}) extremal. Here, we note that $F$ given above can also
be used to describe excitation of the monopole ($\alpha=-1$) and
the quadrupole ($\alpha=2$) modes separately.  For
spherically symmetric trap ($\lambda_{0}=1$) the two modes get decoupled and $\alpha = -1$ and
$\alpha = 2$ correspond to the minimum and the maximum values of the
excitation energies, respectively.

Using the energy functional given by Eq. (\ref{1}) along with the expansion (\ref{2})
we find after a tedious although straightforward  algebra
the following expressions for the moments $m_{1}$ and $m_{3}$:
\begin{equation}
m_{1} = \frac{4\hbar^{2}}{m^{2}} \left (\frac{U_{\bot}}{\left (
\omega_{\bot}^{0}\right )^{2}} + \alpha^{2} \frac{U_{z}}{\left (
\omega_{z}^{0}\right )^{2}}\right ),
\label{9}
\end{equation}
\begin{equation}
m_{3} = \frac{8\hbar^{4}}{m^{2}}\left [T_{\bot} + U_{\bot} + \alpha^{2}\left ( T_{z} + U_{z}
\right ) + \left (1 - \alpha/2\right )^{2}\left ( E_{int}^{1} + \frac{9}{4}E_{int}^{2}\right )
\right ].
\label{10}
\end{equation}
In the above equations $T_{\bot}$ and $U_{\bot}$ ( $T_{z}$ and $U_{z}$ ) are
the transverse components ($z$-component) of the kinetic and potential
energies, respectively. By substituting
Eqs. (\ref{9}) and (\ref{10}) in Eq. (\ref{6}) we get following
expression for the frequency of $m^{\prime}=0$ mode
\begin{equation}
\Omega_{m^{\prime}=0}^{2} = 2\frac{\left [T_{\bot} + U_{\bot} +
\alpha^{2}\left ( T_{z} + U_{z} \right ) + \left (1 -
\alpha/2\right )^{2}\left ( E_{int}^{1} +
\frac{9}{4}E_{int}^{2}\right ) \right ]}
{ \left
(\frac{U_{\bot}}{\left ( \omega_{\bot}^{0}\right )^{2}} +
\alpha^{2} \frac{U_{z}}{\left ( \omega_{z}^{0}\right )^{2}}\right
)}.
 \label{11}
\end{equation}
The frequency of collective oscillations is determined by
substituting the values of energy components from the variational
calculation and  making
$\Omega^{2}_{m^{\prime}=0}$ extremal with respect to $\alpha$.
The variation with respect to $\alpha$ leads to a quadratic
equation in $\alpha$ and the two roots, corresponding to the
maximum ($\alpha_{+}$) and the minimum ($\alpha_{-}$) values of
$\Omega^{2}_{m^{\prime}=0}$, are
\begin{equation}
\alpha_{\pm} = -\frac{B}{2} \pm\frac{1}{2}\sqrt{B^{2} + 4C},
\label{12}
\end{equation}
with
\begin{equation}
B  =  \frac{\left (2f_{2}f_{4} + f_{3}f_{4}/2 - 2f_{1}f_{5} - 2f_{3}f_{5}\right )}{f_{3}f_{5}},
\nonumber 
\end{equation}
and 
\begin{equation}
C  =  \frac{f_{4}}{f_{5}},
\end{equation}
where $f_{1} = T_{\bot} + U_{\bot}$, $ f_{2} = T_{z} + U_{z}$,
$f_{3} = E_{int}^{1} + \frac{9}{4} E_{int}^{2}$, $f_{4} =
U_{\bot}/(\omega_{\bot}^{0})^{2}$ and $f_{5} =
U_{z}/(\omega_{z}^{0})^{2}$. In the next Section we first apply
these results to estimate analytically the correction to the
collective frequency due to the LHY term  within the TF approximation followed by the results of
our variational calculation.
\section{Results and discussion}
\subsection{Analytical estimate within TF approximation}
Before discussing the results of our variational calculation we
present the analytical estimate of collective frequencies within
the TF approximation. We generalize  the perturbative results of
Ref. \cite{pitaevskii,braaten2} and also discuss the validity
regime of these results. For convenience we consider the case of
spherically symmetric trap ($\lambda_{0}=1$). Generalization  for the axially symmetric trap is
straightforward. It is easy to
verify that for spherically symmetric trap $B=-1$ and $C=2$ and
the two roots are $\alpha=2$ and $\alpha=-1$ corresponding to the
quadrupole and the monopole modes, respectively. It is well known
that the frequency of quadrupole mode is not affected by the
LHY term \cite{pitaevskii}.  Therefore we focus our attention on the monopole
mode. By substituting $\alpha=-1$ in Eq. (\ref{11}) the
frequency of the monopole mode of bosons in a spherical trap ($U =
U_{\bot} + U_{z}$ ) within  the TF approximation
($T_{\bot}=T_{z}=0$) can be written as
\begin{equation}
\frac{\Omega^{2}}{\left (\omega_{\bot}^{0}\right )^{2}} = \left
[\frac{U + \frac{9}{4}E_{int}^{1} +
\frac{81}{16}E_{int}^{2}}{U}\right]. \label{13}
\end{equation}
Hereafter  we drop for convenience the subscript $m^{\prime}=0$ in $\Omega$. To
eliminate $U$ from the above equation we make use of the
virial relation \cite{banerjee}
\begin{equation}
2T - 2U + 3E_{int}^{1} + \frac{9}{2}E_{int}^{2} = 0,
\label{virial}
\end{equation}
and obtain within the TF approximation
\begin{equation}
 \frac{\Omega^{2}}{\left (\omega_{\bot}^{0}\right )^{2}} =  \left [\frac{\frac{15}{4}E_{int}^{1} +
\frac{117}{16}E_{int}^{2}}{\frac{3}{2}E_{int}^{1} +
\frac{9}{4}E_{int}^{2}}\right]. \label{14}
\end{equation}
We note that the expression given by Eq. (\ref{14}) is exact
within the TF limit. To derive an analytical expression for the
frequency we express $E_{int}^{1}$ and $E_{int}^{2}$ in terms of
parameter $x=\sqrt{n(0)a^{3}}$. This is done by using the
analytical expressions for the total energy and chemical potential
\cite{dalfovo,banerjee} obtained within the TF approximation with
the LHY term included in the expansion $\epsilon (n)$. These expressions
are:
\begin{eqnarray}
E_{int}^{1} & = & \frac{2}{7}\mu_{TF} - \frac{9}{16}\mu_{TF}x , \nonumber \\
E_{int}^{2} & = &  \frac{5}{8}\mu_{TF}x ,
\label{15}
\end{eqnarray}
where $\mu_{TF}$ is the chemical potential of trapped bosons
obtained within the TF approximation \cite{dalfovo}. It is important to note that the
expressions for $E_{int}^{1}$ and $E_{int}^{2}$ given above (Eq.
(\ref{15}))  are approximate in nature and neglect terms of order
${\cal O}(na^{3})$ and, thus, expected to be valid in the small
gas parameter regime. It has been recently shown \cite{fabrocini}
that the expressions in Eq. (\ref{15}) are not valid for very large gas parameters and lead
to results which differ considerably from the exact results .
In the large-gas-parameter regime one needs
to solve the nonlinear equation of the density more accurately with the help of numerical methods.
Consequently, to compute the frequencies of collective oscillation (for that matter any observable)
in the large-gas-parameter regime within the TF approximation one needs a non-iterative or
non-perturbative method to solve the nonlinear equation. In this paper we take a different route
to achieve this goal. Instead of solving the time-dependent differential equation for the density
(or the time-dependent GP equation) we exploit the sum-rule approach of the many-body response
theory along with the accurate ground-state density determined variationally to compute the
frequencies of collective oscillations. We describe the results of
such a calculation in the next subsection. 

Now by substituting Eq. (\ref{15}) in Eq.
(\ref{14}) we get following expression for the frequency of
the monopole collective mode within the BGP theory:
\begin{equation}
\frac{\Omega}{\omega_{0}} = \left [\frac{5\left (1 +
\frac{147}{64}x\right )}{1 + \frac{21} {16}x}\right ]^{1/2},
\label{16}
\end{equation}
where $\omega_{0}$ is the angular frequency of the spherically
symmetric trap. In the limit of $x\longrightarrow 0$ above
expression (Eq.(\ref{16})) correctly reduces to the mean-field
result $\Omega_{mf}=\sqrt{5}\omega_{0}$ within the TF approximation
\cite{stringari}.

It is customary to define the fractional change in the
frequency as \cite{pitaevskii,braaten2},
\begin{equation}
\delta = \frac{1}{2}\left (\frac{\Omega^{2} - \Omega_{mf}^{2}}
{\Omega_{mf}^{2}}\right ).
\end{equation}
By using Eq. (\ref{16}) we get
\begin{equation}
\delta = \frac{\frac{63}{128}x}{1 + \frac{21}{16}x}. \label{17}
\end{equation}
In the limit of $x<<1$ we recover the results of Refs.
\cite{pitaevskii,braaten2}. Fig. 1 shows the fractional
change $\delta$ as a function of dimensionless parameter $a/a_{0}$
(where $a_{0}$ is the Bohr radius of hydrogen atom) for $N=10^{4}$
$^{85}$Rb atoms trapped in a spherical trap with
$\omega_{0}/2\pi=\left (\left (\omega_{\bot}^{0}\right
)^{2}\omega_{z}^{0}\right)^{1/3}/2\pi=12.83Hz$, where
$\omega_{\bot}^{0}=2\pi\times 17.5$Hz and
$\omega_{z}^{0}=2\pi\times 6.9$Hz \cite{cornish} .  
The values of the parameter $a/a_{0}$ are chosen such that the condition for the TF approximation,
that is , $Na/a_{h0}>>1$ is satisfied.
Moreover, as discussed above Eq.(\ref{15}) (and hence Eq. (\ref{16})) is valid only for small gas
parameter values and accordingly we have restricted ourselves to $a/a_{0} = 5000$. 
It is evident from Fig.1  that the correction to the monopole frequency due to inclusion of LHY
term in the interatomic correlation energy obtained by us is
slightly lower than the corresponding results of Refs.
 \cite{pitaevskii,braaten2}. In particular for $a/a_{0}=5000$ we
obtain $7\%$ change in the monopole frequency as compared to the
correction of $9\%$ predicted in Ref. \cite{pitaevskii,braaten2}.
It is only for very small values of $x$ (less than $10^{-3}$) that
fractional corrections obtained in Ref. \cite{pitaevskii,braaten2} match with Eq. (\ref{17}).

\subsection{Variational results}
Before discussing the results for the excitation frequencies obtained by variational method,
we first demonstrate that this method is capable of yielding sufficiently accurate results for the 
ground-state properties of BEC even for large gas parameter values, as achieved in the 
experiment of Cornish et al. \cite{cornish}.
The  parameters used in this calculation are similar to the ones used in the previous section.
We present these results in Table I and II for the spherically symmetric and the axially symmetric traps
respectively. To check the accuracy, we compare our variational results with those of
Ref. \cite{fabrocini} (presented in parenthesis in Table I and II) obtained by numerically solving
the modified GP equation. The results
in Table I and II clearly show that the variational method gives quite accurate results for the
ground-state properties as they compare very well with the numbers of Ref. \cite{fabrocini}.
Therefore the variational method is indeed suitable for describing the ground-state properties
even in the regime of large gas parameters.
The accuracy of our variational results was further checked by verifying that the generalized
virial relation as given in Eq. (\ref{virial}) is satisfied. In all the
variational calculations the virial relation was satisfied up to the fifth decimal place or better.
This indicates that not only the total energy but its different components are also obtained
accurately by the variational method, which is quite crucial for the calculation of the frequencies
of collective oscillations.

Now we focus our attention on the results for the frequencies of collective 
oscillations of trapped bosons. Figs. 2 and 3, show
the upper ($\Omega_{u}$) and the lower ($\Omega_{l}$) frequencies respectively, associated with
the breathing mode as a function of the dimensionless parameter $a/a_{0}$. The results presented in 
Figs. 2 and 3 correspond to the axially symmetric trap which is more relevant from the 
experimental point of view.
Notice that the maximum value 
of the parameter $a/a_{0}$ is chosen to be $10^{4}$ in accordance with the experimental number
\cite{cornish} and the corresponding value of the peak
gas parameter is $\approx 10^{-2}$. For comparison we also show in these
figures the corresponding results obtained within the GP theory.  
It is important to note here that the GP results shown in Figs. 2 and 3 are also obtained 
variationally by dropping the LHY term from the expansion (\ref{2}) and thus, retaining only the
first term. The results corresponding to the GP case obtained by us match quite
well with those already reported in the literature 
\cite{stringari,kimura,edwards,singh,esry,sinha,dalfovoex,you,hutchinson}.
It can be clearly seen from
Fig. 2 that for small values of the gas parameter  the numbers obtained
within the GP and the BGP theory are quite close. The difference between
these two results, however, increases with $a$. For example, for
$a/a_{0}=8000$ the BGP number for $\Omega_{u}$ is $6\%$ higher than the corresponding mean-field
GP result. Thus we conclude that the introduction of LHY term in the interatomic correlation 
energy results in a significant correction to the upper frequency $\Omega_{u}$ of 
$m^{\prime}=0$ mode of the collective oscillations.
Considering the accuracy of current experiments \cite{stamperkurn} in measuring the frequencies of
the collective excitations, it should be possible to observe these corrections experimentally.
It is also important to note that the frequency $\Omega_{u}$ obtained within the GP theory
remains almost constant with the increase in the parameter $a/a_{0}$. In contrast, 
the  BGP theory  predicts $\Omega_{u}$ growing monotonically with $a/a_{0}$. Therefore 
a signature of the predictions of the BGP theory can be observed by measuring $a/a_{0}$ dependence
of the excitation frequency  $\Omega_{u}$.

Next we focus our attention on the lower frequency $\Omega_{l}$ of the breathing mode which 
is shown in Fig. 3.
For small values of the gas parameter the frequency of the lower branch obtained via GP theory
is very close to the corresponding BGP theory result. In the same regime of the gas parameter
the frequency $\Omega_{l}$ decreases with  the gas parameter both for the GP and
the BGP cases. The decrease
of the lower frequency with an increase in the gas parameter value has already been theoretically
predicted \cite{stringari,edwards,esry,you} and experimentally verified as well \cite{jin}.
However, in contrast to the GP case,  frequency of the lower branch for the BGP case
starts increasing, albeit slowly, after a critical value of $a/a_{0}\approx 2200$.
It is also noteworthy that the change in the frequency $\Omega_{l}$ obtained from the BGP theory is considerably 
lower than that in the upper frequency $\Omega_{u}$.
For example,  for $N=10^{4}$ and $a/a_{0}=8000$ the change in the lower frequency is just $0.8\%$ 
while that in the upper frequency is $6\%$.

We conclude this section by making a comparison of  the frequencies obtained
by using the analytical expression as
give in  Eq. (\ref{16}) with the results obtained by the variational approach. The comparison is
shown in Fig. 4 where we plot the frequency of the monopole mode of a condensate in a 
spherically symmetric trap as a
function of $a/a_{0}$ obtained using Eq. (\ref{16}) (dashed line) and using the
variational method (solid line). The range of values of $a/a_{0}$ is chosen such that the  
TF approximation (i.e. $Na/a_{h0}>>1$ ) is satisfied. Accordingly the minimum value of 
$a/a_{0}$ in Fig. 4 is chosen to be $10^{3}$ which
corresponds to $Na/a_{h0}=174$.  It can be seen  that the results obtained using Eq.(\ref{16}) are systematically
higher than the corresponding results from variational calculation and the difference between 
the two grows as the
value the gas parameter is increased. Moreover, we also note that the two results are reasonably
close for values of $a/a_{0}$ of the order of 3500 ($n(0)a^{3}\approx 3.77\times 10^{-3}$). 
Therefore, we emphasize that the correction obtained within the TF approximation as given by Eq. (\ref{16}) is
valid as long as the gas parameter is of the order of $10^{-3}$. As discussed earlier, for 
condensates with the values of the gas parameter
beyond $10^{-3}$ one needs to evaluate $E_{int}^{1}$ and $E_{int}^{2}$ in a non-perturbative manner.
The variational approach employed in this paper accomplishes exactly this task.

\section{Conclusion}
In this paper we have calculated the frequencies
of collective excitation of a trapped condensates in the large-gas-parameter regime using 
the BGP theory . To go beyond the
GP theory we have included a higher-order term (LHY term) in the interatomic
correlation energy obtained from the ground-state energy of a uniform Bose gas.
The calculations of the frequencies were performed by employing
the sum-rule approach of many-body response theory along with the variational method
to obtain the ground-state energy components. We have also derived an analytical expression for 
the corrections to the frequencies of the compressional mode valid for small gas parameters
within the TF approximation. The results
of our variational calculations clearly show that for large gas parameter values, 
the frequencies of the compressional mode characterized by the index $m^{\prime}=0$ are
significantly altered.  
Although the change in the lower branch of $m^{\prime}=0$ mode, $\Omega_{l}$, is
small, we observe a new qualitative feature. 
The frequency $\Omega_{l}$ increases slowly as the gas parameter exceeds  
a critical value in contrast to a monotonically decrease predicted within the GP theory.
It would be very interesting to verify the predictions of the BGP theory by measuring collective
excitation frequencies for  condensates with large gas 
parameter as reported in Ref. \cite{cornish}.

\section*{Acknowledgement}
It is a pleasure to thank Dr. S. C. Mehendale for helpful discussions and critical reading of the
manuscript.

\newpage
\begin{table}
  \centering
\caption{Results for the ground-state properties of $^{85}$Rb atoms trapped in an isotropic trap
with $\omega_{\bot}^{0}/2\pi=17.5Hz$. Total energy $E_{1}/N$ and chemical potential $\mu_{1}$
are in units of $\hbar\omega_{\bot}^{0}$. Numbers in parenthesis are results
of Ref. \cite{fabrocini}. The values of parameter $Na/a_{h0}$ given in the second column signify
the importance of the interaction energy  over the kinetic energy.}
\tabcolsep=0.2in
\begin{center}
\begin{tabular}{|c|c|c|c|}
\hline
$a/a_{0}$ &$Na/a_{h0}$ & E$_{1}/N$ & $\mu_{1}$ \\
\hline
1400 & 243.38 & 10.03 & 14.05 \\
     & & (10.01) & (13.97) \\
3000 & 522.0 & 14.14 & 19.92  \\
     & &  (14.10)& (19.84) \\
8000 & 1392.0 & 23.52 & 33.54   \\
     &  & (23.43) & (33.38) \\
10000 & 1740 &  26.75  & 38.24  \\
      & &  (26.63) & (38.06) \\
\hline
\end{tabular}
\end{center}
\end{table}

\begin{table}
  \centering
\caption{Results for the ground-state properties of $^{85}$Rb atoms trapped in axially symmetric
 trap with $\omega_{\bot}^{0}/2\pi=17.5Hz$ and $\lambda_{0}=0.39$. Total energy $E_{1}/N$ and
 chemical potential $\mu_{1}$ are in units of $\hbar\omega_{\bot}^{0}$. Numbers in parenthesis are results
of Ref. \cite{fabrocini}. The values of parameter $Na/a_{h0}$ given in the second column signify
the importance of the interaction energy  over the kinetic energy.}
\tabcolsep=0.2in
\begin{center}
\begin{tabular}{|c|c|c|c|}
\hline
$a/a_{0}$ & $Na/a_{h0}$ & E$_{1}/N$ & $\mu_{1}$ \\
\hline
1400 & 284.25 &7.36 & 10.25 \\
     & &  (7.33) & (10.22) \\
3000 & 609.0 & 10.35 & 14.55  \\
     & & (10.31)& (14.51) \\
8000 & 1624.0 & 17.18 & 24.49   \\
     & & (17.09) & (24.38) \\
10000 & 2030.0 &  19.53  & 27.92  \\
      & & (19.43) & (27.79) \\
\hline
\end{tabular}
\end{center}
\end{table}

\clearpage
\newpage

\begin{figure*}
\caption{Fractional shift of the frequency of $m^{\prime} =0$ mode in a spherically symmetric trap
with $\omega_{0} = 2\pi\times 12.83 Hz$. The solid line corresponds to Eq. (\ref{17}) and the
dashed line shows the results of Refs. \cite{pitaevskii,braaten2}.}
\end{figure*}

\begin{figure*}
\caption{The frequencies (in units of $\omega_{\bot}^{0}$ ) of the upper branch of the collective mode $m^{\prime}=0$ of $10^{4}$ 
$^{85}$Rb atoms confined in an axially symmetric trap with $\lambda_{0}=0.39$ and 
$\omega_{\bot}^{0}=2\pi\times 17.5 Hz$. The solid line represents results for the BGP case while 
the corresponding 
GP results are shown by the dashed line. }
\end{figure*}

\begin{figure*}
\caption{The frequencies (in units of $\omega_{\bot}^{0}$ ) of the lower branch of collective mode
 $m^{\prime}=0$. The parameters are same as in Fig. 2}
\end{figure*}

\begin{figure*}
\caption{The frequencies (in units of $\omega_{0}$ ) of the monopole mode of $10^{4}$ $^{85}$Rb atoms
confined in a spherically symmetric trap with parameters same as Fig. 1. The solid line
represents results for the BGP case and the dashed line shows results obtained from Eq. (\ref{16})}.
\end{figure*}
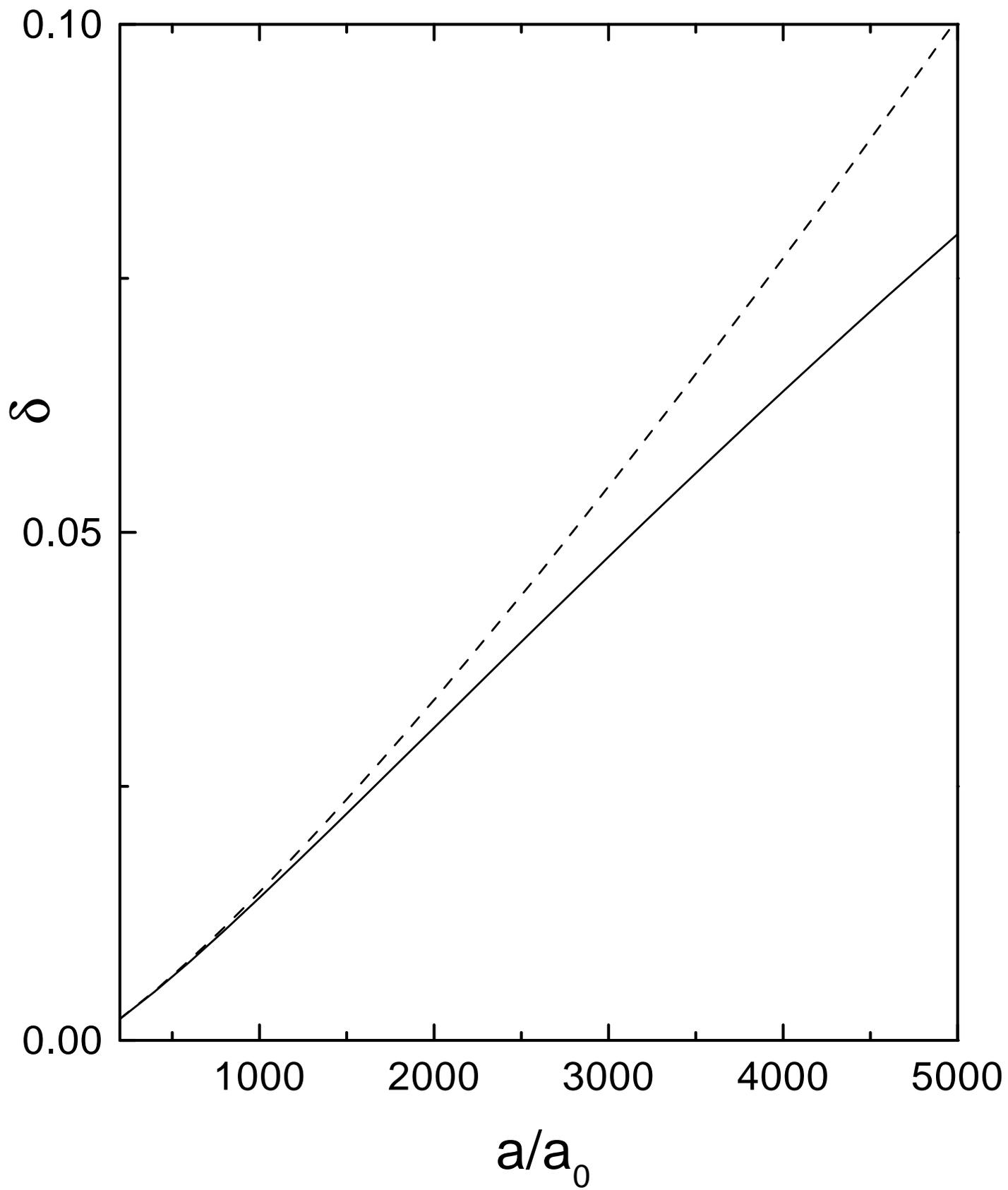
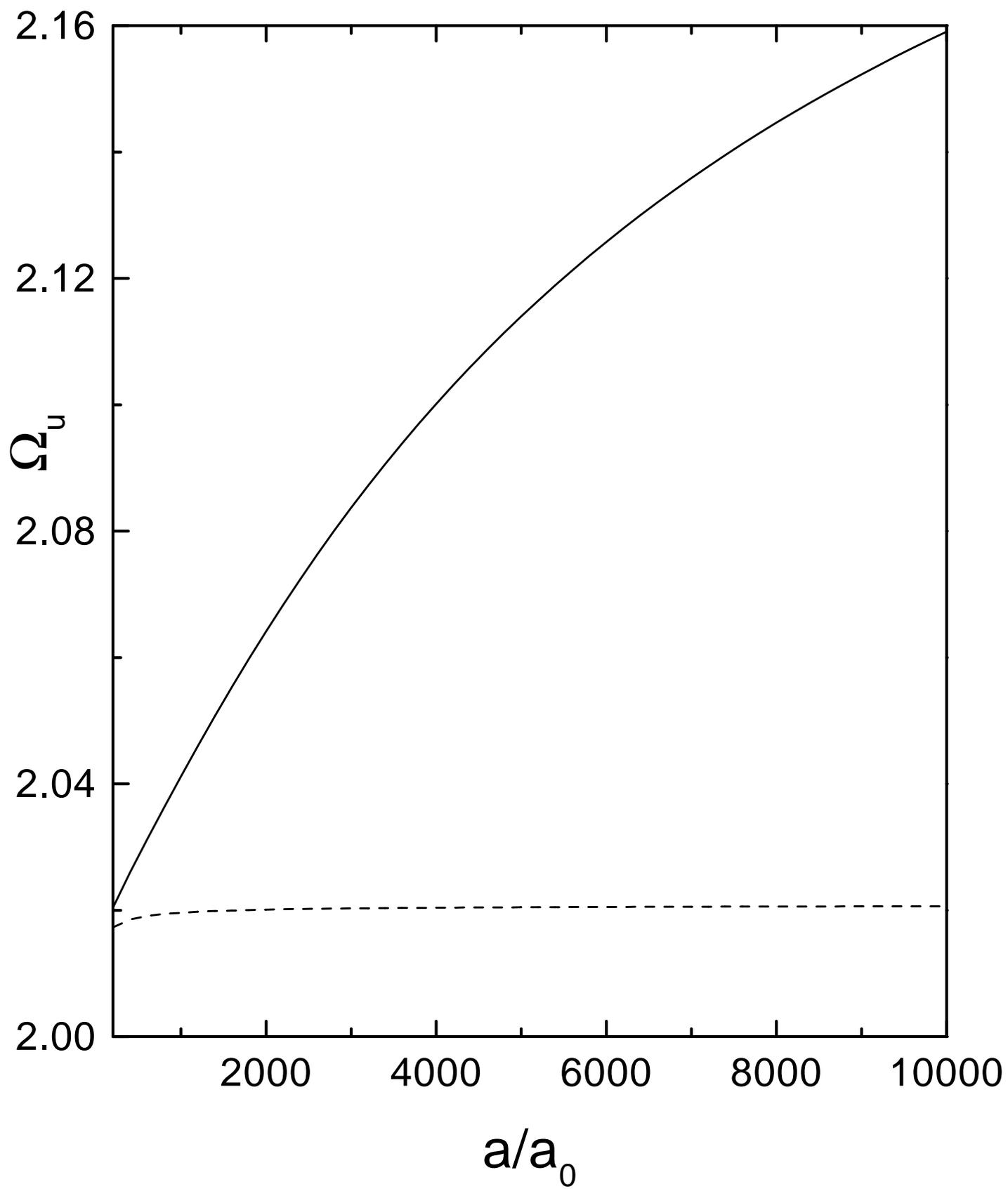
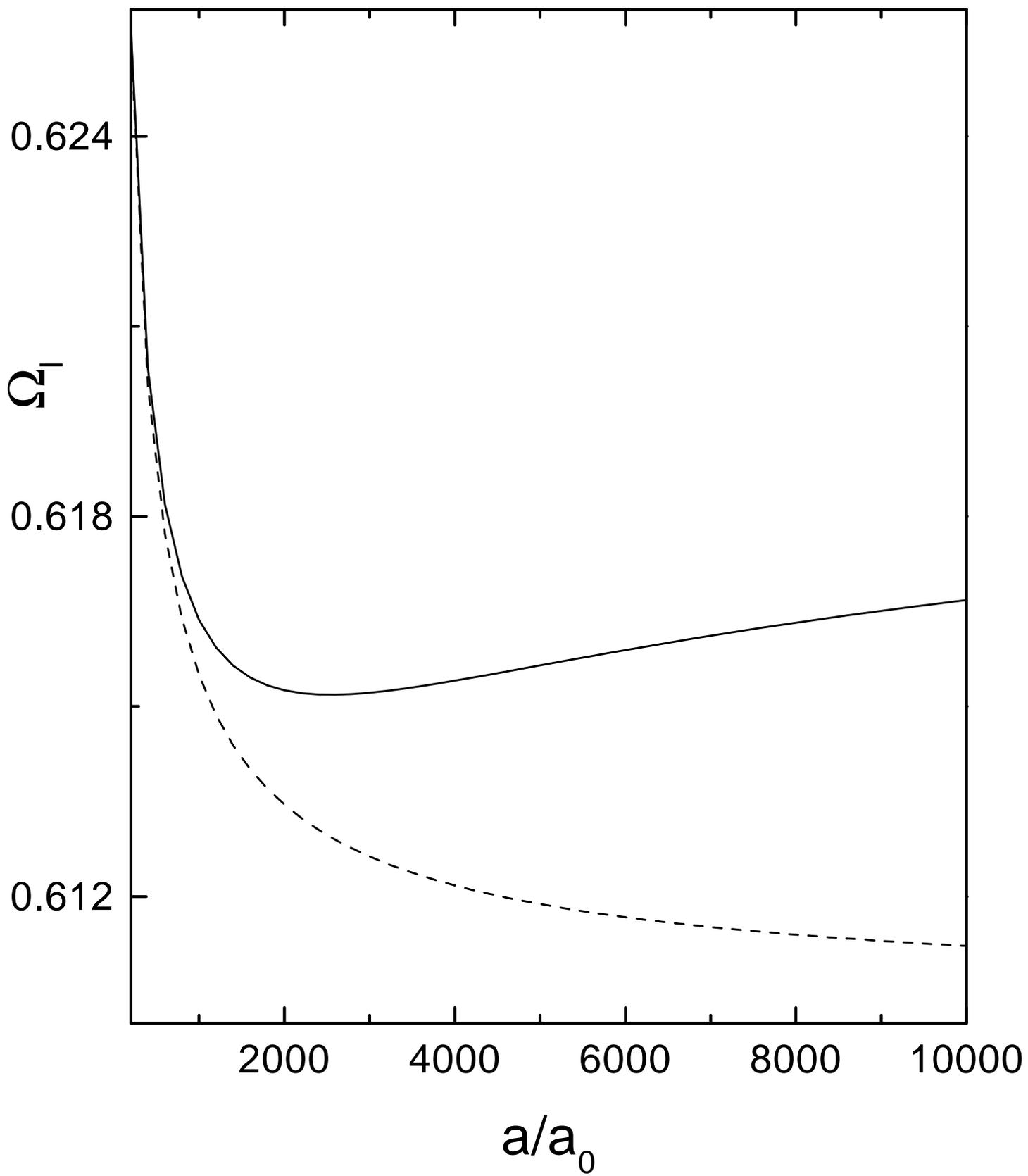
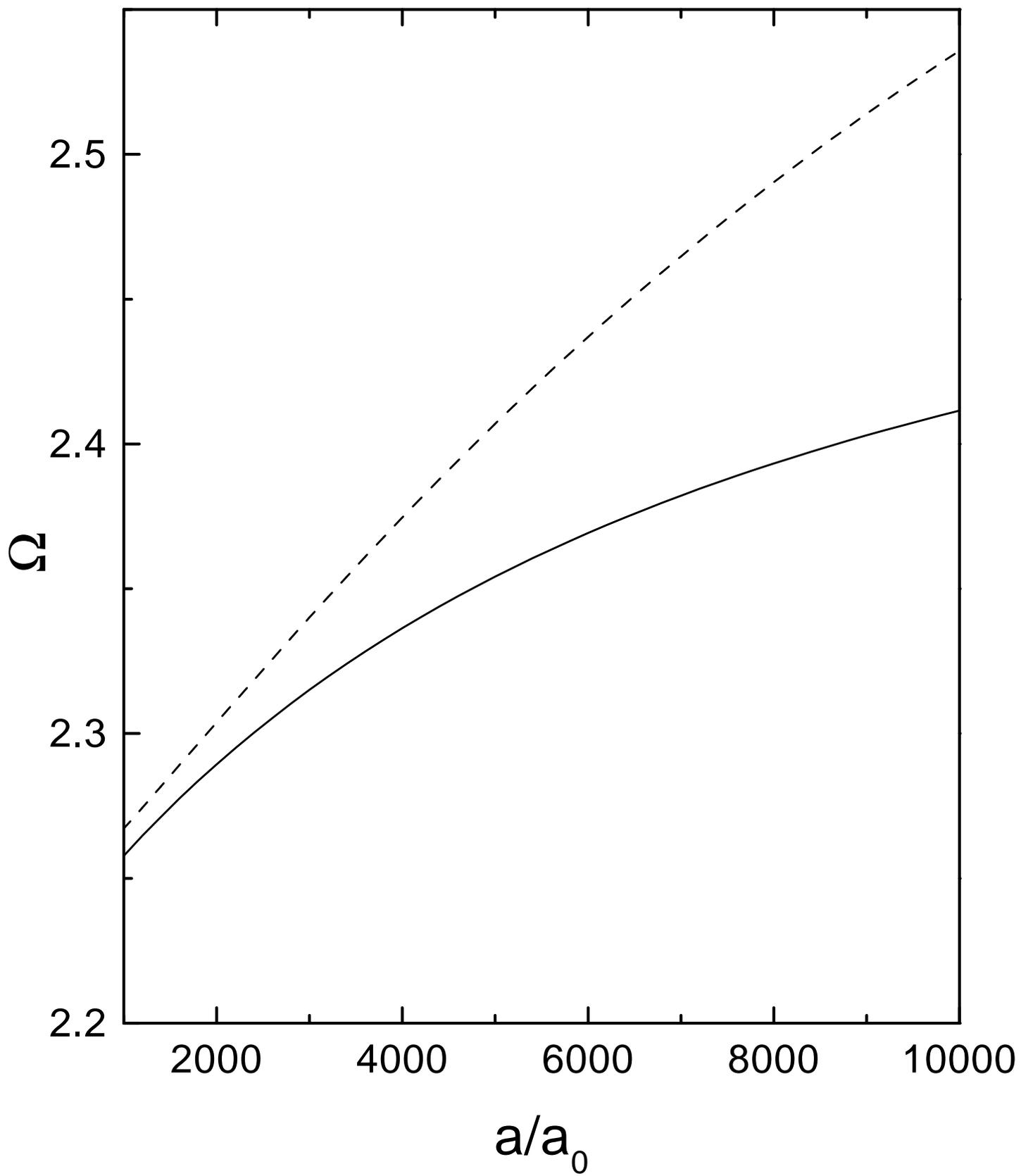
\end{document}